\documentclass[letter,traditabstract]{aa} 

\usepackage{graphicx}
\usepackage{orcidlink}
\usepackage{txfonts}
\hypersetup{pdfborder={0 0 0}}
\hypersetup{colorlinks=true, linkcolor=blue, citecolor=blue, urlcolor=blue}

\begin{document}

   \title{The mass-metallicity relation at $z\gtrsim 3$ down to $M_{\star}\simeq 10^4~M_{\sun}$}

   \subtitle{A local perspective using the metallicity distribution of RR Lyrae stars}

   \author{M. Bellazzini \orcidlink{0000-0001-8200-810X}
\inst{1}
          \and
          T. Muraveva\orcidlink{0000-0002-0969-1915}\inst{1}
          \and
          A. Garofalo\orcidlink{0000-0002-5907-0375}
          \inst{1}}

   \institute{INAF - Osservatorio di Astrofisica e Scienza dello Spazio di Bologna, via Gobetti 93/3, 40127 Bologna, Italy\\
              \email{michele.bellazzini@inaf.it}
       }

   \date{Received xxxxx; accepted xxxx}

 \abstract{The mass-metallicity relation (MZR) is a fundamental scale law of galaxies. It
is observed to evolve with redshift in unresolved galaxies up to
$z> 6$. However, observational constraints limits our view at such early epochs to galaxies with $M_{\star}\gtrsim 10^7~M_{\sun}$. On the other hand, in the local Universe the
MZR can be traced down to the faintest end of the galaxy luminosity
function ($M_{\star}\simeq 10^2~M_{\sun}$) but we have access only to its
present-day realization. We propose to use RR
Lyrae stars to get the mean metallicity of local dwarf galaxies at the
early epoch  in which these variable stars were formed ($z\gtrsim3$), opening a new window on the evolution of the MZR across cosmic times down to the lowest mass. We use available data for a sample of Milky Way satellites to show
that  indeed the evolution of the MZR from the epoch of the formation of RR
Lyrae to the present day can be traced with this approach, with results broadly compatible with those inferred from high $z$ galaxies from nebular emission lines.
The limitations of the approach as well as possible ways to refine the analysis are also briefly discussed.
}

\titlerunning{The mass-metallicity relation at $z\gtrsim 3$ down to $M_{\star}\simeq 10^4~M_{\sun}$.}
   \maketitle
   
%

\section{Introduction} \label{sec:intro}
  The mass-metallicity relation (MZR), 
  the correlation between the stellar mass of a galaxy  ($M_{\star}$) and its mean abundance of chemical elements heavier than Helium, is a fundamental scale law of galaxies, holding either
when nebular metallicity (measuring the current level of chemical enrichment of the inter stellar medium (ISM)), or the stellar metallicity, (somehow integrating over the galaxy star formation history (SFH)), are considered \citep[see][MM19 hereafter, and references therein]{maio19}. The MZR is believed to originate from the complex interplay between star formation, pollution of the ISM by metals from dying stars and retention of enriched gas by the galaxy potential well, outflow and inflow of gas due to SN and AGN feedback and accretion of gas from the environment, respectively
\citep[MM19;][]{baker23,looser24}. 

\begin{figure*}[!th]
    \centering
    \includegraphics[width=0.9\textwidth]{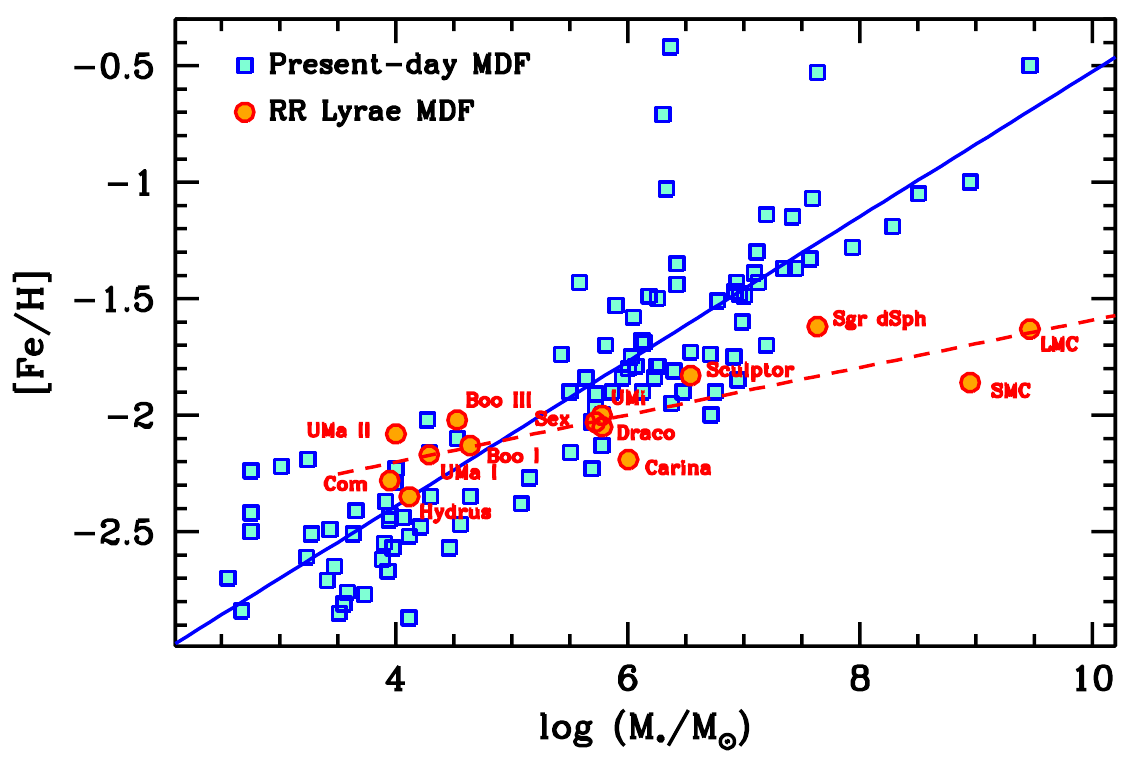}
    \caption{MZR for local dwarf galaxies with metallicity from the present-day MDF and from the RR Lyrae MDF, probing the $z \gtrsim 3$ epoch. The straight lines are linear fits to each MZR.}{\label{fig:mmr}}   
\end{figure*}

The MZR is observed to evolve with cosmic time: on one side the zero-point metallicity (normalisation) increases from early epochs to the present-day, due to the cumulative effect of subsequent generations of stars on the overall chemical enrichment of the galaxy \citep[MM19;]{maio08,curti23}, on the other side the slope of the log~$M_{\star}$ - log~Z\footnote{Where Z is the ratio of the mass in elements heavier then He to the total mass of baryons. See MM19 for the definition of the various parameters used to express the metallicity of stars and galaxies.} seems to flatten towards early epochs (high redshift $z$), likely reflecting significant differences in the effect of feedback in different mass ranges (\citealt{curti24,sarkar25,fujimoto25}, but see \citealt{rowland25}, for different results).
While the advent of the James Webb Space Telescope (JWST) has significantly moved the lower mass limit to which the relation can be traced in the $z>3$ regime down to $M\gtrsim 10^7~M_{\sun}$ \citep[][]{nakajima23,curti23,curti24,cherme24, chakra24,sarkar25,rowland25}, we are still orders of magnitude above the range reached in the local Universe, where the stellar MZR is traced, in the ultra-faint dwarf galaxy regime, down to $\gtrsim 10^2~M_{\sun}$ \citep[][P24, hereafter]{mc12,pace24}. 

Therefore, by looking at the integrated light of distant galaxies in different ranges of redshift we have access to the MZR at different epochs, back to the 
dawn of the Universe ($> 11.5$~Gyr ago for $z>3.0$ and $> 13.0$~Gyr ago for $z>7.0$, when the age of the Universe was less than 2~Gyr and 1~Gyr, respectively; \citealt{planck}) but we cannot reach the faintest end of the galaxy luminosity function. On the other hand, with the sample of local galaxies that can be resolved into individual stars we can sample the present-day (pd) MZR down to $M_{\star} \simeq 10^2~M_{\sun}$, but we do not have access to earlier epochs.

Here we propose a way to circumvent the latter limitation, at least partially. RR Lyrae variables are generally recognised as very old stars \citep[age$\gtrsim 10.0$ Gyr, see][and references therein]{catelan04,savino20,clementini23}. They are typical of globular clusters, where they are usually older than 12.5~Gyr \citep{dotter10,vande16} and they are found in most local dwarf galaxies. Assuming that the chemical evolution of dwarf galaxies proceeds following a monotonic, approximately single-valued age-metallicity relation \citep[AMR, see, e.g.,][]{pagel98,monte98,ac19}, it can be considered that RR Lyrae sample a narrow age window of the AMR at an early epoch and their metallicity distribution function (MDF) records a snapshot of the chemical enrichment of the galaxy at that time. Hence, we can use the mean of the RR Lyrae MDF, when available, to obtain a glance at the MZR at early epochs (ee), that is for $z\gtrsim 2$ if age$\gtrsim 10.0$~Gyr, or $z\gtrsim 5$, if the age of RR Lyrae in dwarf galaxies is similar to that of RR Lyrae in globular clusters. Here we take $z\gtrsim 3$ as a reference value. Then we can compare the ee-MZR to the pd-MZR to see if we can trace the evolution of the MZR through cosmic epochs also by using information in the local Universe and down to very low masses. Potentially, we can also explore if the view of the MZR we get from old stars in local galaxies is compatible with what we get from high-z galaxies. Below, we will show and briefly discuss these comparisons. 


\section{The local stellar MZR at early epochs} \label{sec:mmr}

To produce the stellar pd-MZR we took stellar masses and spectroscopic mean metallicities (in terms of iron abundance [Fe/H]) for local dwarf galaxies from P24. In the large majority of cases, the mean metallicities by P24 are obtained from MDFs of Red Giant Branch stars, thus sampling the epochs from $\sim 13$~Gyr to $\simeq 2$~Gyr ago. They should be considered as the result of the integration of the chemical evolution up to its current status. The present-day stellar masses, on the other hand, are obtained by P24\footnote{See {\tt https://local-volume-database.readthedocs.io}} from the integrated absolute magnitude of the galaxies in the V band ($M_V$) by adopting $M/L_V=2$, a simple and widely used approach \citep[see, e.g.,][]{mc12}.

For the RR Lyrae MDF we proceeded as follows. We adopted the homogeneous set of photometric metallicities derived by \citet{muraveva25} from the pulsation period and Fourier decomposition parameters of the light curves of RR Lyrae stars provided in {\it Gaia} Data Release 3 (DR3, \citealt{clementini23}). Then, we matched this catalogue with lists of RR Lyrae belonging to nearby dwarf galaxies from the literature. In this way, we assembled samples of RR Lyrae with metallicity for fourteen dwarf galaxies, all satellites of the Milky Way (MW), 
keeping all dwarfs for which at least two RR~Lyrae members with metallicity estimates from \citet{muraveva25} were found in the {\em Gaia} DR3 catalogue and not including candidate extra-tidal members \citep{vivas20}.  The resulting MDF are shown in Appendix~\ref{app:mdf}, Fig.~\ref{fig:mdf}. The sample sizes and the literature references adopted for each galaxy are listed in Table~\ref{tab:rrmet}, in the same Appendix.
It is important to note that the very small samples for the Ultra Faint Dwarfs \citep[UFD;][]{belo13,simon19} Coma~Berenices, Bootes~I, Bootes~III, Hydrus~I, Ursa Major~I and Ursa Major~II, in spite of their scantiness, can be considered as representative of the metallicity at the epoch of RR Lyrae formation for these galaxies, since they virtually include their entire population of such variables, or a significant fraction of it \citep{vivas20}. The faintest dwarfs in our sample with RR~Lyrae MDF have $M_{\star}\simeq 10^4~M_{\sun}$; for lower stellar masses the total number of evolved stars is so low that it is unlikely to include one RR Lyrae star, let alone two. 
From each sample we derived the weighted mean [Fe/H] and we adopt it as the galaxy metallicity at the epoch of RR Lyrae formation (see Tab.~\ref{tab:rrmet}).

While the mean of the RR Lyrae MDF gives us a window on the chemical enrichment of the galaxies at early epochs ($z\gtrsim 3$) we do not have a simple way to recover the second key ingredient to build the ee-MZR, that is the stellar mass at the same epoch. However, for the purpose of the detection of the effect of galaxy chemical evolution on the MZR the present-day mass is probably an acceptable proxy. Indeed, the SFH of nearby dwarfs suggests that most of them built up $>50\%$ of their stellar mass more than 10~Gyr ago \citep{weisz14}, hence the present-day stellar mass should be just within a factor of 2-3 from the mass at that epoch. On the other hand, some of them may have lost significant fraction of their total mass by tidal interaction with the MW. However, this process may be much more impactful for the dark matter halo of these galaxies than for their stellar component. For example, in the models of the disruption of the Sgr~dSph galaxy by \citet{vasi20}, successfully reproducing the observed properties of the system, during the process of tidal disruption up to the present day, the dark mass of Sgr is reduced by a factor $\simeq 8$, while the stellar mass by a factor $\la 2$.
Finally, for the purpose of ascertaining that the MZR was already in place for local galaxies at early epoch and that it has evolved since then to its present-day form, the ranking in stellar mass may be more important than the actual value, and this should have been broadly preserved through evolution (e.g., the SMC has always been significantly more massive than Draco, etc.). 

The comparison between the pd-MZR and the ee-MZR for local dwarfs is displayed in Fig.~\ref{fig:mmr}. The evolution that occurred between the epoch of RR Lyrae formation and the present-day is evident for the most massive dwarfs ($M_{\star}\gtrsim 10^6~M_{\sun}$), that were able to form stars for longer periods of time (up to the present epoch for the LMC and SMC) and/or were likely more efficient in producing metals and in retaining chemical enriched gas ejected by SNe. In these galaxies the mean metallicity at $z\gtrsim 3$ was significantly lower than today.
At lower masses the mean metallicity of the galaxies at the two epochs becomes indistinguishable, within the uncertainties, as we reach the mass regime where feedback from the first generation of stars suddenly quenched the SF and
even re-ionisation may have played a role in the pristine interruption the chemical evolution \citep{weisz14,aparicio16,curti24}. In practice Fig.~1 can be considered as a first glance to the mass regime where the cosmic evolution of the MZR begins to become perceivable. The MZR seems to be already in place at the epoch of RR Lyrae formation, the correlation coefficient is $\rho=0.837$ and the rms=0.11~dex, to be compared to $\rho=0.870$ and rms=0.28~dex for the pd-MZR. The evolution is traced by the steepening of the slope of the relation, from $\frac{d[Fe/H]}{dlogM}=0.10\pm 0.02$ for the ee-MZR to $\frac{d[Fe/H]}{dlogM}=0.31\pm 0.02$ for the pd-MZR. 

To get a quantitative assessment of the reliability of a correlation based only on 14 points, several of which are clustered around $M_{\star}\simeq10^4~M_{\sun}$ and $M_{\star}\simeq10^6~M_{\sun}$,
we computed the Spearman's rank correlation coefficient for the ee-MZR, obtaining $\rho_S=0.767$. The probability to get a $\rho_S$ value equal or larger than this by mere chance is $P_S=0.0014$. Hence, the statistical significance of the ee-MZR shown in Fig.~\ref{fig:mmr} is high but not  ironclad (see below).

\begin{figure}[!th]
    \centering
    \includegraphics[width=\hsize]{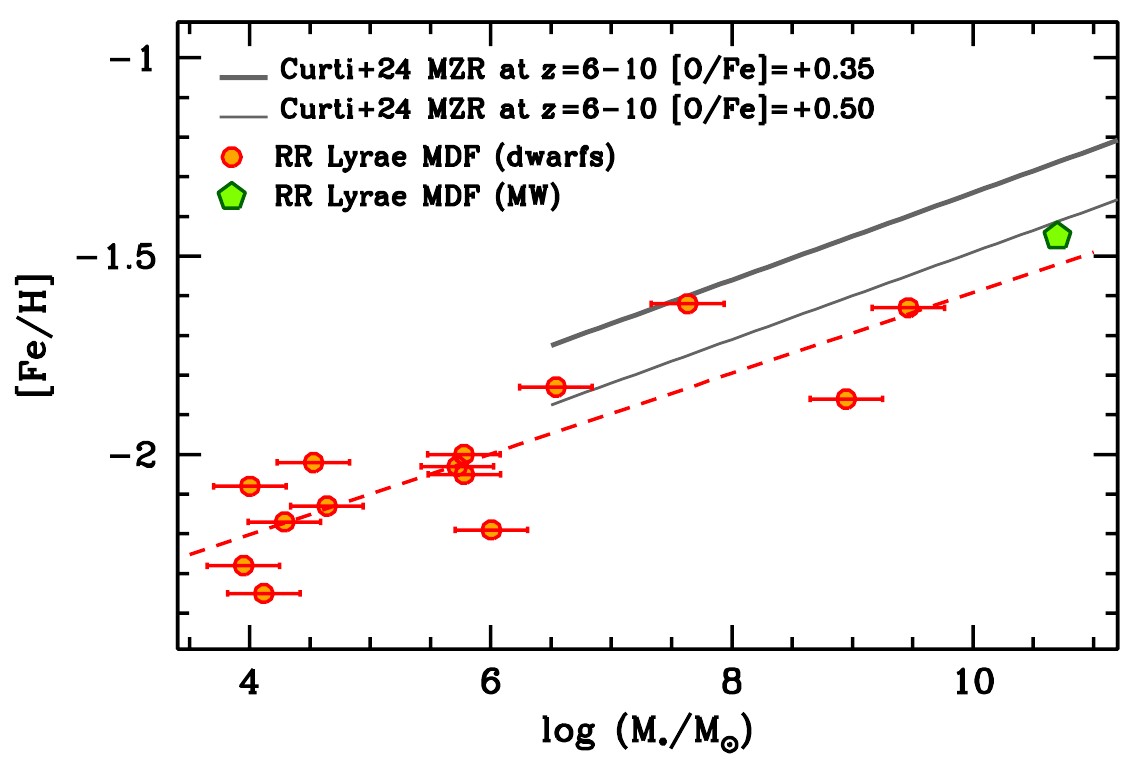}
    \caption{MZR for local dwarf galaxies with metallicity from the RR Lyrae MDF (same symbols as in Fig.~\ref{fig:mmr}; a factor 2 uncertainty in mass is assumed, for reference) compared to the linear fit to the MZR of $z=6-10$ galaxies by \citet{curti24} (grey solid lines), assuming two different values of [O/Fe] enhancement. The dark green pentagon filled in pale green shows the location of the MW galaxy.{\label{fig:mmr_hz}}}
\end{figure}

In Fig.~\ref{fig:mmr_hz} we show that also the MW appears to follow to the same ee-MZR defined by local dwarfs. Here the stellar mass of the MW has been taken from \citet{bh16} and the metallicity is the mean metallicity of the RR Lyrae in the \citet{muraveva25} sample after excluding all the stars attributed to the satellites\footnote{To account for the discovery of a significant population of metal-rich ([Fe/H]$>-1.0$) RR Lyrae stars in the Bulge and in the Thin Disc of the MW, the hypothesis that such population is largely composed by stars younger than 10~Gyr living in binary systems has been proposed \citep[see][and references therein]{bobrick24}. 
This population should not have any significant impact in the present analysis since RR Lyrae with [Fe/H]$>-1.0$ are extremely rare or absent in the dwarf galaxies, and constitute only the $\simeq 13\%$ of the RR Lyrae attributed to the MW in our sample. Hence, the RR Lyrae MDFs considered here should be dominated by genuinely old stars.}. The MW is likely the galaxy whose $M_{\star}$ had the most significant change since $z\gtrsim 3$ \citep{marinacci14}, among those included in Fig.~\ref{fig:mmr_hz}, hence its location in this plot must be considered with particular caution. Moreover, since we are considering only one additional point, the match may be fortuitous.   On the other hand, it may suggest that the ee-MZR can be traced with the RR Lyrae MDF also beyond the stellar mass range of dwarfs, and independently of the galaxy type. It may be interesting to note that adding the MW to the dwarfs the Spearman's coefficient of the ee-MZR reaches $\rho_S=0.811$, with $P_S=0.00025$.

In the same plot we compare the local ee-MZR inferred here with the MZR inferred by \citet{curti24} from a sample of galaxies at high redshift ($6\le z\le 10$), observed with JWST. The metallicity of the \citet{curti24} MZR has been converted from 12+log(O/H) into [Fe/H] with the solar oxygen abundance by \citet{dela06} and correcting for the $\alpha$-enhancement with respect to the solar abundance pattern that is typically observed in dwarf galaxies in this low metallicity regime \citep[see, e.g.,][]{kirby11}. In particular we adopted [O/Fe]=$+$0.35, according to the O abundances by \citet[][]{hasselquist21}, and [O/Fe]=$+$0.50, more similar to the average $\alpha$-enhancement observed by \citet[][]{kirby11}.

First of all, it seems remarkable that the local ee-MZR and the high-z MZR have very similar slopes, both significantly different from their present-day/low-z counterparts. Second, the relatively small difference in normalisation (0.25~dex and 0.1~dex for the two [O/Fe] cases considered) can be, at least partly, attributed to the inhomogeneity of the two metallicity scales. For instance, the measures by \citet{curti24} are based on the analysis of nebular spectra, while the metallicity by \citet{muraveva25} is calibrated on stellar spectroscopy. Moreover, it is important to note that the lower values of the stellar mass of dwarfs expected at early epochs would also help to improve the match between the two MZRs, reducing the difference in the normalisation.
The key point of Fig.~\ref{fig:mmr_hz} is that the local ee-MZR and the high-z MZR appear strikingly similar, also considering the large galaxy-to-galaxy scatter observed in the high-z MZR \citep{curti24}. 

\section{Conclusions}
\label{sec:conc}

The simple experiment presented here is intended to demonstrate that the MDF of RR~Lyrae may be used to investigate the MZR at early epochs in the local Universe, down to the faintest end of the galaxy luminosity function. The analysis can be refined by using available cumulative SFH and dynamical modelling of the past interaction of each dwarf with the MW to infer more precisely the stellar mass at the epoch of RR~Lyrae formation, by taking into account the role of SFH in the build-up of the MZR, that is, in fact, a projection of a more general mass-SF-metallicity relation (MM19), by including more galaxies and by increasing the size of the RR~Lyrae samples,  by fixing, within the uncertainties, the $[O/Fe]$ ratio for each galaxies from direct spectroscopic measures in RR~Lyrae or in RGB stars at the same metallicity within the same galaxy, by enhancing the homogeneity between the metallicity scales used in the pd-MZR and ee-MZR (and with the high-z MZR), by deriving MDFs of RR Lyrae from spectroscopic measures on significant samples of individual stars, etc. All these possible refinements of the analysis are clearly beyond the range and scope of the this Letter. In general, the precision of the quantitative conclusions that it would be possible to draw with the proposed technique will depend and rely on the reliability of models and independent measures that will be used to make the input data for local galaxies as homogeneous as possible with those available for their high-z counterparts, e.g., by following the paths listed above. Moreover, at present, it does not seem possible to overcome the limitations inherent to the uncertainty in the actual age of the RR~Lyrae and to the fact that an uncertainty of 1~Gyr in the hypothesised age, at such old ages, implies a large uncertainty in the corresponding z (e.g., going from $z\simeq 3$ to $z\simeq 5$ for age changing from 11.6~Gyr to 12.6~Gyr). 
In any case, the adopted approach may provide a new tool to investigate the early days of the galaxies in our neighbourhood.
In particular, if the match between the local ee-MZR and the high-z MZR suggested in Fig.~\ref{fig:mmr_hz} is confirmed with better data and more detailed analysis, it would provide a major consistency test for two complementary views the evolution of baryonic matter in galaxies, as the same fundamental scaling law would be found by two completely independent ``time-machines'', that is peering into the distant past of the Universe by looking at high redshift galaxies or by looking at very old stellar populations in nearby systems.

\begin{acknowledgements}
We are very grateful to Roberto Maiolino for reading the original manuscript 
and for his useful comments and suggestions. MB acknowledge the support to this study by the INAF Mini Grant 2023 (Ob.Fu. 1.05.23.04.02 – CUP C33C23000960005) CHAM – Chemo-dynamics of the Accreted Halo of the Milky Way (P.I.: M. Bellazzini). A.G. acknowledges the support to this study that has been provided by the Agenzia Spaziale Italiana (ASI) through contract ASI 2018-24-HH.0 and its Addendum 2018-24-HH.1-2022.
This work made use of data from the European Space Agency
(ESA) mission Gaia (\url{https://www.cosmos.esa.int/gaia}), processed
by the Gaia Data Processing and Analysis Consortium (DPAC; \url{https://www.cosmos.esa.int/web/gaia/dpac/consortium}). Funding for
the DPAC has been provided by national institutions, in particular,
the institutions participating in the Gaia Multilateral Agreement.

\end{acknowledgements}

%
  \bibliographystyle{aa} 
   \bibliography{aa54043-25} 
%

\begin{appendix}

\section{RR Lyrae MDFs}
\label{app:mdf}

\begin{figure}[!th]
    \centering
    \includegraphics[width=\columnwidth]{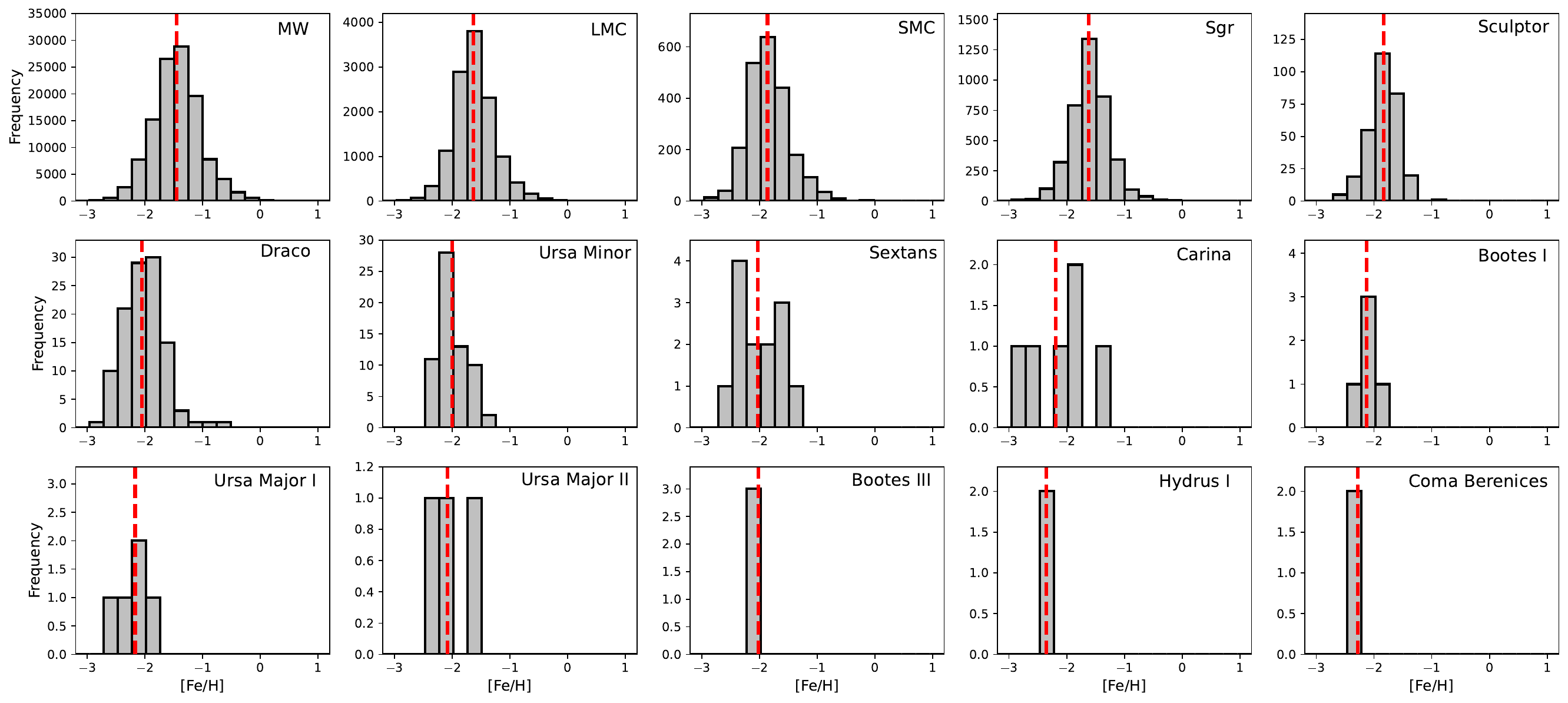}
    \caption{RR Lyrae MDF for the dwarfs in our sample and for the MW (see Tab.~\ref{tab:rrmet}). The red dashed lines mark the location of the weighted mean of the MDFs.}{\label{fig:mdf}}   
\end{figure}

In Fig.~\ref{fig:mdf} we show the RR~Lyrae MDFs for all the dwarf galaxies considered in this work and for the MW. The weighted mean metallicity and standard deviation, the sample size and the references for the membership of the RR~Lyrae for each galaxy are listed in Table~\ref{tab:rrmet}.

\begin{table}[!th]
\centering
\caption{Mean metallicity from RR~Lyrae MDF \label{tab:rrmet}}             
\begin{tabular}{l c c r l}        
\hline\hline                 
Gal name & $\langle [Fe/H]\rangle$& $\sigma_{[Fe/H]}$ & $N_{\star}$ & references\\    
&[dex] & [dex] &  & \\ 
\hline                        
MW & $-$1.45 & 0.40 & 115828 & this work\\
LMC& $-$1.63 & 0.34 & 12239 & \citet{cusano21}\\
SMC& $-$1.86 & 0.33 &  2203 & \citealt{muraveva18} \\
Sgr~dSph & $-$1.62 & 0.31 & 3950 & \citet{ramos22} \\
Sculptor   & $-$1.83 & 0.24  & 297 & \citet{martinez15}\\
Draco      & $-$2.05 & 0.34  & 112 & \citet{muraveva20} \\
Ursa~Minor & $-$2.00 & 0.25  & 64 & \citet{garofalo24} \\
Sextans    & $-$2.03 & 0.36  & 13 & \citet{vivas19} \\
Carina     & $-$2.19 & 0.40  &  6 & \citet{dallora03} \\
Bootes~I   & $-$2.13 & 0.16  &  5 & \citet{dallora06,siegel06,vivas20}\\
Ursa~Major~I & $-$2.17 &  0.27 & 5 & \citet{garofalo13} \\
Ursa~Major~II & $-$2.08 &  0.26 & 3 & \citet{dallora12,vivas16,vivas20}\\
Bootes~III & $-$2.02 & 0.01 & 3 & \citet{sesar14,vivas20}\\
Hydrus~I   & $-$2.35 & 0.07 & 2 & \citet{vivas20} \\
Coma~Berenices & $-$2.28 & 0.02 & 2& \citet{musella09,vivas20}\\
\hline                                   
\end{tabular}
\tablefoot{Weighted means. The typical uncertainties on individual 
metallicity in the \citet{muraveva25} is $\gtrsim 0.4$~dex. 
The Sgr~dSph sample includes both stars in the main body and in the tidal tails
of the galaxy.}
\end{table}
%

\end{appendix}

\end{document}